\def\degr{$^{\circ}$}

\newcommand\g{\ensuremath{\gamma}}%

\documentclass[a4paper]{article}

\usepackage{icrc2013}

\title{Study of the Very High Energy emission from Supernova Remnants with H.E.S.S}

\shorttitle{ICRC 2013 Template}

\authors{
Diane Fernandez$^{1}$,
Joachim Hahn$^{2}$,
Vincent Marandon$^{2}$,
Matthieu Renaud$^{1}$ ,
Aion Viana$^{2}$  for the 
H.E.S.S. collaboration.
}

\afiliations{
$^1$ Laboratoire Univers et Particules de Montpellier, Montpellier, France \\
$^2$ Max-Planck-Institut f\"ur Kernphysik, Heidelberg, Germany \\
}

\email{diane.fernandez@lupm.univ-montp2.fr; joachim.hahn@mpi-hd.mpg.de}

\abstract{ From radio and higher-frequency observations, more than 300 SNRs have 
been discovered in the Milky Way, of which 220 fall into the H.E.S.S. 
Galactic Plane Survey. However only 50 SNRs are coincident with a 
H.E.S.S source and in 8 cases the VHE emission is firmly associated with 
the SNR. The H.E.S.S. dataset includes now more than 8 years of 
observations and it is of great interest to extract VHE flux upper 
limits from undetected SNRs. These new measurements can then be used to 
test the standard paradigm of the SNRs as the origin of Galactic cosmic 
rays.
In this contribution, the H.E.S.S. results on the population of SNRs and 
the subsequent constraints on the cosmic-ray acceleration efficiency in 
these sources will be presented.}


\begin{document}
\maketitle

\section{Introduction}

  Since the 1930s' \cite{bz}, supernova remnants (SNRs) remain the most likely sources of Galactic cosmic rays (CRs). According to the diffusive shock acceleration (DSA) and magnetic field amplification mechanisms, particles can be accelerated up to the knee at the SNR shock surface. Nonetheless despite several decades of multi-wavelength observations no compelling {\it direct} evidence in favor of efficient CR acceleration in SNRs has been observed yet. 
  
  The High Energy Stereoscopic System (H.E.S.S.) is an array of four imaging atmospheric Cherenkov telescopes (IACTs) situated in Namibia. It covers a field of view (FOV) of 5\degr\, and detects \g-rays above an energy threshold of $\sim$\,100\,GeV. The primary particle direction and energy are reconstructed with an energy resolution of $\sim$\,15\% and an angular resolution of $\sim$\,0.1\degr. 
 
About 50 H.E.S.S. sources are coincident with a Galactic SNR, among which 8 are firmly associated with the SNR. This value includes 5 shell emissions (Vela Jr \cite{velaH}, HESS J1731--347 \cite{j1731H}, RCW 86 \cite{rcwH}, RX J1713.7--3946 \cite{j1713H}, SN 1006 \cite{sn1006H}) and 3 SNR interacting with a molecular cloud (MC) (W51C \cite{w51H,w51M}, W49B \cite{w49H}, W28 \cite{w28H}). About 10 other H.E.S.S. sources are possibly associated with a SNR shell or SNR-MC emission. The remaining sources are either associated with a pulsar wind nebula (PWN) emission or without any clear counterpart. 
  
  Using the large $\sim$\,8 years dataset of H.E.S.S., we investigate the VHE emission from radio and/or X-ray SNRs listed in the GreenÕs and University of Manitoba's catalogues and continue previous studies on VHE emission from SNRs that are not detected with H.E.S.S.\cite{Anne}. 

In order to test the standard paradigm of the SNRs as the origin of Galactic cosmic rays, a safe sample of SNRs is selected: this sample includes $\sim$\,100 SNRs showing no hint of VHE signal. The source selection is detailed in Section~\ref{sec-selec}. For each SNR of this sample an upper limit (UL) on the VHE \g-ray flux in the energy band between 1\,TeV and 100\,TeV is extracted. Assuming a hadronic emission a constraint on the SNR explosion energy going into proton acceleration is calculated. Since an additional leptonic component might be present in the VHE flux, the former hypothesis is conservative and allows to constrain the acceleration efficiency of protons trapped within the SNR shock surface. Our results are presented in Sec.\ref{sec-results}.
  

\section{Source Selection and analysis}
\subsection{Source Selection}
\label{sec-selec}
The investigated data set is made up by the H.E.S.S. galactic plane survey (HGPS), conducted between the years 2004-2012. In total, it consists of over 2800 hours of observation with a sensitivity of better than $\sim3\%$ of the Crab in the innermost galactic region \cite{HGPS}.

We use the SNRcat \cite{SNRCAT} catalogue to determine our test positions. This catalogue lists SNRs as they are observed from radio to very high energies. The astrophysical sources include shell-like, composite and centre-filled SNRs. It amounts to a total of $\sim$300 SNRs.

Of these $\sim$300 SNRs, about 220 fall into the HGPS region. However, there are many H.E.S.S. sources in this region \cite{HGPS}, so source confusion is an issue. This is especially problematic since for a later comparison with SNR VHE emission models, one has to limit the contribution of signal of other sources (and especially non SNRs) into the analysis region. 

\begin{figure}[h!!!]
  \centering
  \includegraphics[width=0.5\textwidth]{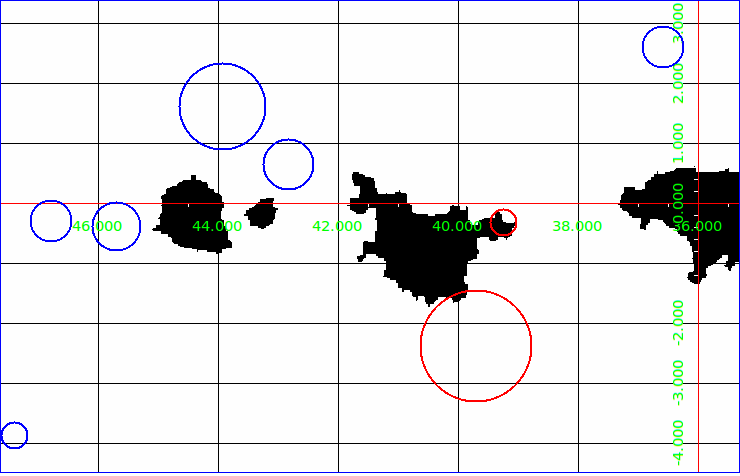}
  \caption{Source selection: Blacklisted pixels are shown in black. Analysis regions are represented by circles. If analysis regions encompass blacklisted pixels (red), the corresponding source is removed from the sample.}
  \label{deselect}
 \end{figure}

The HGPS significance map \cite{HGPS}, derived with the HAP-TMVA analysis and \textit{standard cuts} \cite{TMVA} and correlated with a radius of 0.2$^\circ$, is used to identify contingent regions of signal around known H.E.S.S. sources. To that end, the positions of known H.E.S.S. sources are used as seeds around which an iteration is performed: map pixels with a significance of larger than a certain sigma value ($\sigma_T$) are blacklisted for source selection. An example for the resulting regions that are black-listed can be seen in figure \ref{deselect}. 

If the analysis region (see section \ref{sec-analysis}) for a given source encompasses such pixels, the particular source is removed from the source sample.
Also, we excluded from the samples the uncertain SNRs such as G007.5-01.7 or G354.1+00.1. Two sets of sources are selected:
\begin{itemize}
  \item a loose sample, where $\sigma_T = 4$ (125 sources)
  \item a conservative sample, where $\sigma_T = 2$ (96 sources)
\end{itemize}

The observation live time of the analyzed regions span a range from about half an hour to more than 130h with a mean value of about 40h.


\subsection{Analysis}
\label{sec-analysis}


  For each considered SNR the signal extraction region is taken to be a uniform disk of radius $\theta = R_{SNR} + \theta_{PSF} $ , where $R_{SNR}$ is the angular radius of the SNR obtained from radio/X-ray observations. If minor and major axes are provided in the SNRcat, we define the source radius as $R_{SNR} = \sqrt{R_{min} \cdot R_{maj}}$. The H.E.S.S. point spread function (PSF), conservatively approximated by $\theta_{PSF}$ = 0.1\degr \, is added to ensure that the full SNR emission is contained in the ON-region.
  
  Data are analysed with the multivariate analysis method applying \textit{standard cuts} as described in \cite{TMVA}. The presented results are obtained from HGPS count and acceptance maps \cite{HGPS} and are cross-checked with independent calibration chain and the Model Analysis \cite{Modpp} which confirmed the results.

 Assuming a power-law spectrum $\frac{dN}{dE} \propto (\frac{E}{1\rm TeV})^{-\Gamma}$ in the 1-100 TeV energy range, one can estimate the amount of expected excess counts for the assumed spectral shape according to the on-live-time \textit{T} and the acceptance \textit{A(E)} of the selected dataset:

\[
N = \int_{1\rm TeV}^{100 \rm TeV}  T\cdot A(E)\cdot  \Bigg(\frac{E}{1\mathrm{TeV}}\Bigg)^{-\Gamma} \mathrm dE 
\]

 The upper limit $n^{UL}_{99\%}$ at 99\% confidence level (CL) on the amount of excess events coming from the ON-region is derived according to the Rolke$\&$al., 2005 \cite{Rolke} approach.

 The flux upper limits above 1TeV at 99\% CL are then extracted in the following way: 
\[
\Phi^{UL}_{99\%} (>1 \rm TeV) = \frac{n^{UL}_{99\%}}{N} \int_{1\rm TeV}^{100\rm TeV} \Bigg(\frac{E}{1 \rm TeV}\Bigg)^{-\Gamma}\, \mathrm dE 
\]
The DSA mechanism predicts a spectral index of $\sim$\,2 for the CRs accelerated at SNR shock wave. The propagation of CRs in the galactic disk depends on the diffusion coefficient $D_{ISM}(E) \propto E^{\delta}$ with $\delta$ = 0.3...0.6.  and leads to a Galactic CR spectrum at the Earth with an index of $\sim$\,2.75. On the other hand the spectral analysis of individual sources gives an index steeper than 2 in most of the cases. Hence, to standardize the analysis of the SNR sample, we adopt generic values of 2.1 and 2.3 for the spectral index.


\section{Results and Discussion}
\label{sec-results}
  
\begin{figure}[h!!!]
  \centering
    \includegraphics[width=0.5\textwidth]{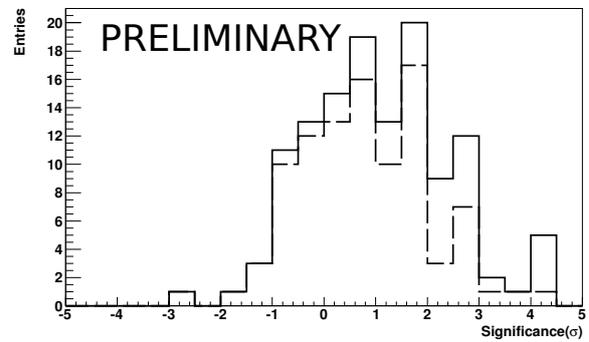}
  \caption{Significance distribution of the SNRcat sample after selection. Solid line: Loose sample, dashed line: conservative sample.}
  \label{sign_dist}
\end{figure}

Figure \ref{sign_dist} shows the significance distributions for the loose (solid line) and the conservative source samples (dashed line). No significant sources are found. 



\begin{figure}[h!!!]
  \centering
    \includegraphics[width=0.5\textwidth]{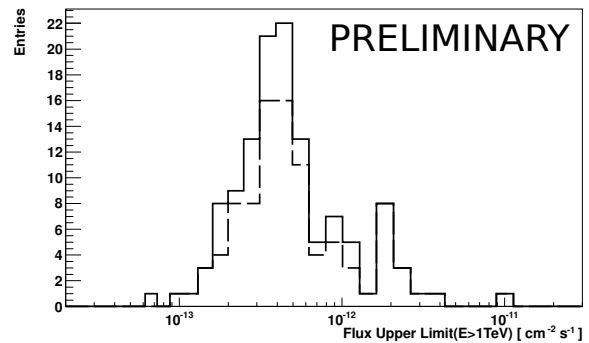}
  \caption{Flux ULs above 1TeV (at the 99\% CL). Solid line: Loose sample, dashed line: conservative sample.}
  \label{ULdist}
\end{figure}

The distribution of the resulting Flux-UL (at the 99\% CL and assuming a spectral index of 2.1) above 1TeV can be seen in figure \ref{ULdist}. It is peaked at around $4\cdot 10^{-13}$cm$^{-2}$s$ ^{-1} $($\sim 2\%$ Crab) which reflects the sensitivity of the H.E.S.S. observations in the inner galactic plane region. 

The obtained UL values allow us to probe the popular hadronic VHE emission model by Drury et al.\cite{DAV} via $\pi_0$-decay. In this model, the gamma ray flux above an energy E can be written as
\[
F_{\gamma} (>E) \approx \frac{1}{4\pi d^{2}} \,\theta \, E_{SN}\, n \,q_{\gamma}  \,E^{1-\alpha} 
\]
where $\theta$ is the CR acceleration efficiency, $n$ is the ambient density around the SNR, $E_{SN}$ is the total kinetic energy output of the SNR and $d$ is the distance to the object. The parent particle energy distribution in the remnant is assumed to be power-law shaped with a spectral index of $\alpha$ over the whole energy range. Finally, the parameter $q$ represents the gamma ray production rate which depends on $\alpha$. 

\begin{figure}[h!!!]
      \centering
        \includegraphics[width=0.5\textwidth]{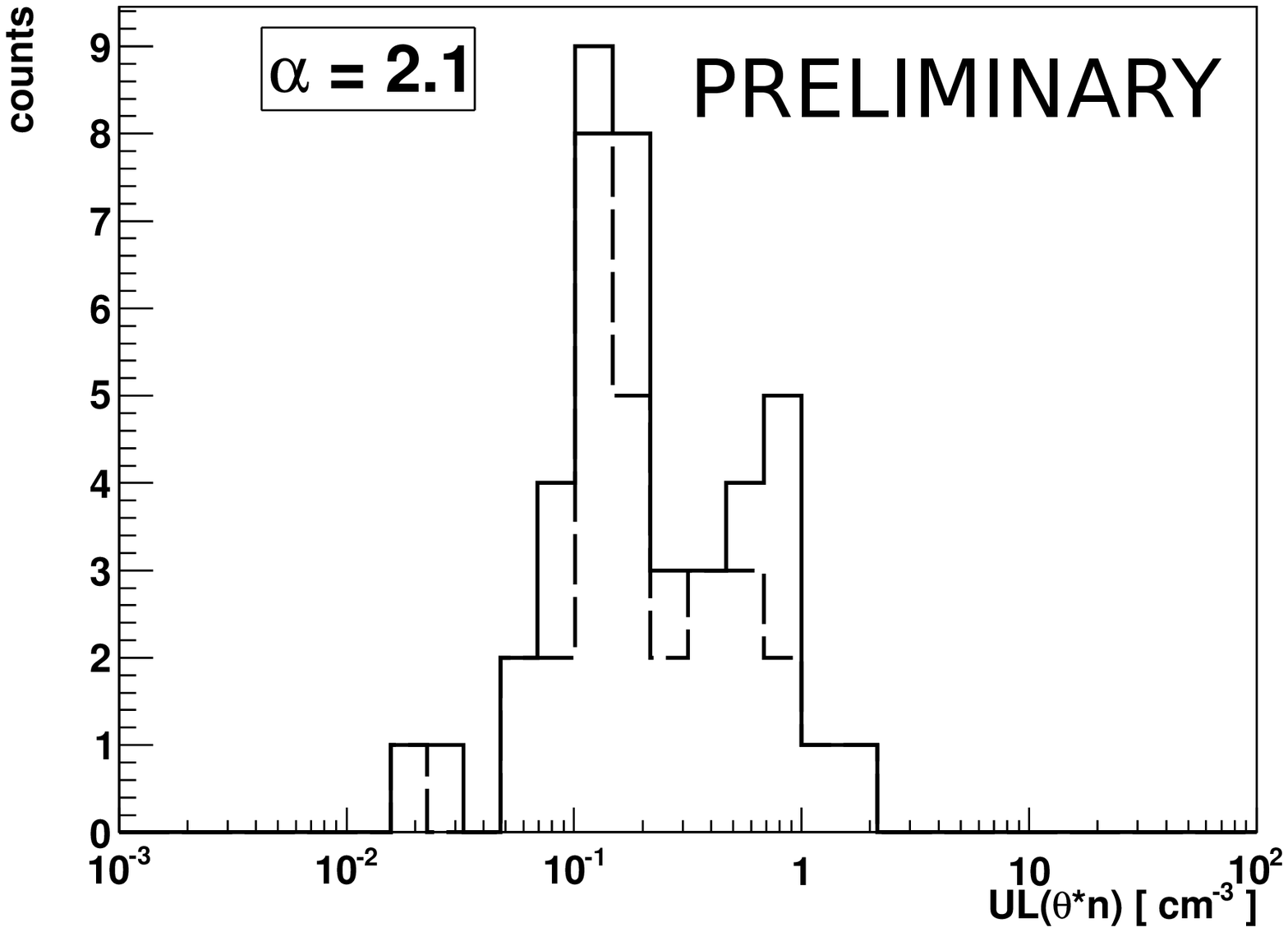}
        \includegraphics[width=0.5\textwidth]{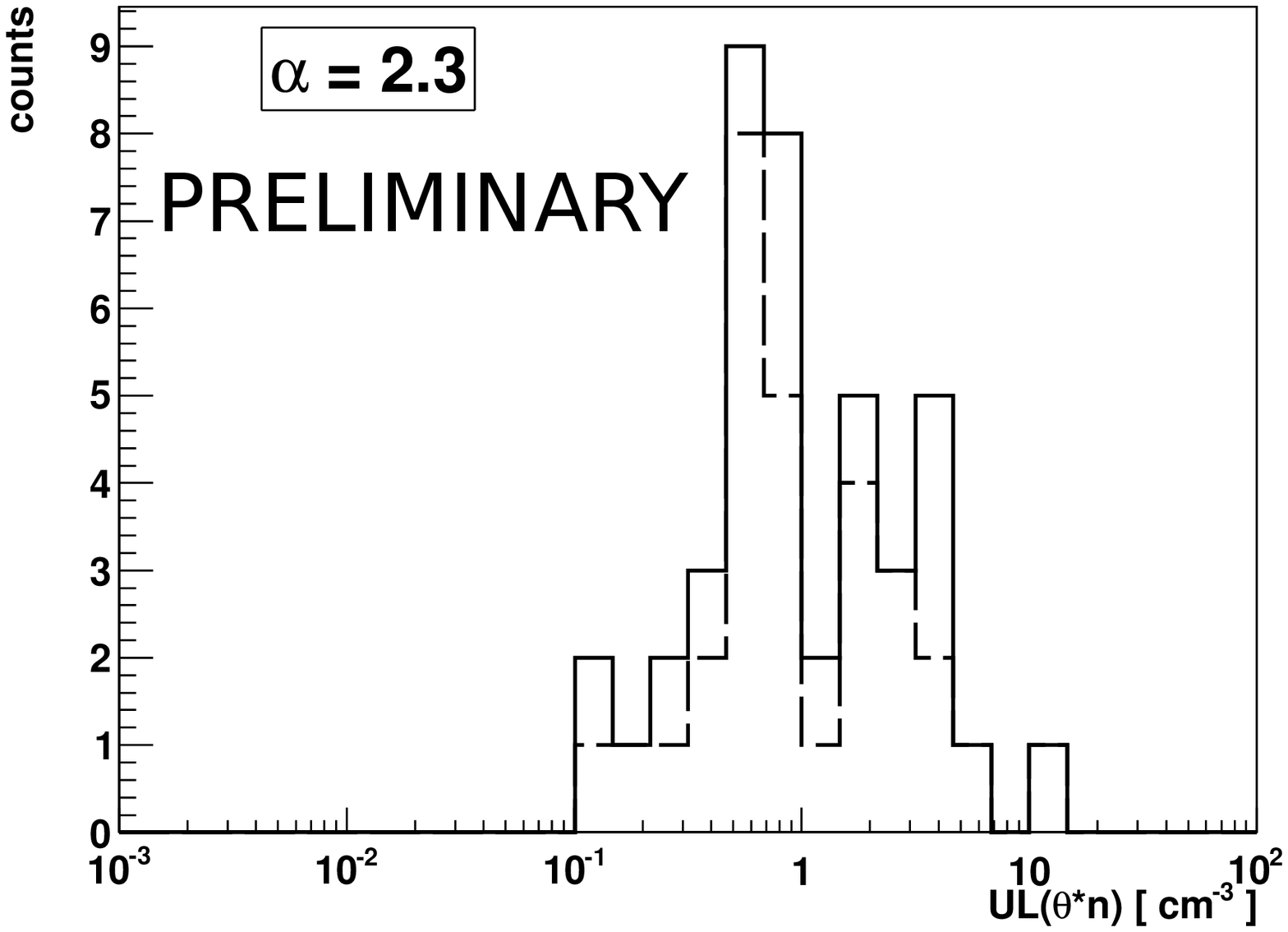}
      \caption{Upper limits on the product $\theta \cdot n$ for assumed parent particle distribution spectral indices of $\alpha = 2.1$ (top) and $\alpha = 2.3$ (bottom). Solid lines: All SNRs with distance estimate (42 sources), dashed lines: SNRs with additional age estimate (30 sources). The sources are taken from the loose sample (see text). }
      \label{constraint}
    \end{figure}

By restricting the SNR sample to sources where a distance estimate is given and assuming a blast energy of $E_{SN} = 10^{51}$erg, it is possible to derive an UL on the product $n\cdot \theta$. The distribution of these constraints is shown in Fig \ref{constraint} for assumed spectral indices of $\alpha = 2.1$ (top) and $\alpha = 2.3$ (bottom).

For an assumed spectral index of $\alpha = 2.1$ the distribution median value is $\langle n\cdot \theta \rangle = 1.65 \cdot 10^{-1}$ ($\langle n\cdot \theta \rangle_{Age} = 1.59 \cdot 10^{-1}$ for sources with additional age estimate) and $\langle n\cdot \theta \rangle = 7.46 \cdot 10^{-1}$ $\left(\langle n\cdot \theta \rangle_{Age} = 7.15 \cdot 10^{-1}\right)$ if $\alpha = 2.3$ is assumed. Assuming typical values of $n$ in the order of 1cm$^{-3}$ puts the result in the expected range for the acceleration efficiency $\theta$ of a few times 0.1. However, there are several limitations to this constraint:
\begin{itemize}
  \item the $\theta$ value is a result of decaying proton in the energy range that corresponds to resulting gamma rays with energies $E>$1TeV and holds only under the assumption of a simple power law over the \textit{whole} proton energy range;
  \item in the 1-100 TeV energy range there might be a spectral cut-off or break in the proton spectrum which would not be accounted for. Indeed, the escape of CRs from the SNR shock surface and/or the interaction of the SNR with a molecular cloud (MC) can significantly change the spectral shape from the one expected with the DSA. Note that a spectral index softer than 2.3 would shift the constraint $n\cdot \theta$ to higher values;
  \item there is no reason to believe that the CR acceleration efficiency $\theta$ is constant with time. CRs could be accelerated to the highest energies at the very beginning of the SNR evolution and the CR acceleration efficiency could decrease significantly after the Sedov-Taylor transition time;
  \item the ambient gas density around SNRs is not very well known and large variations in this value are possible especially for core-collapse (cc) SNRs. A young cc-SNR expanding in the wind of its progenitor might encounter densities of $n\geq10$cm$^{-3}$. At a later phase, due to a steep radial circumstellar gas density profile, the same object could expand into a very thin medium ($n\leq0.1$cm$^{-3}$);
  \item the SNR blast energy E$_{SN}$ can easily vary on the order of a few, especially for core-collapse SNe.
  
\end{itemize}

  \begin{figure}[h!!!]
      \centering
        \includegraphics[width=0.55\textwidth]{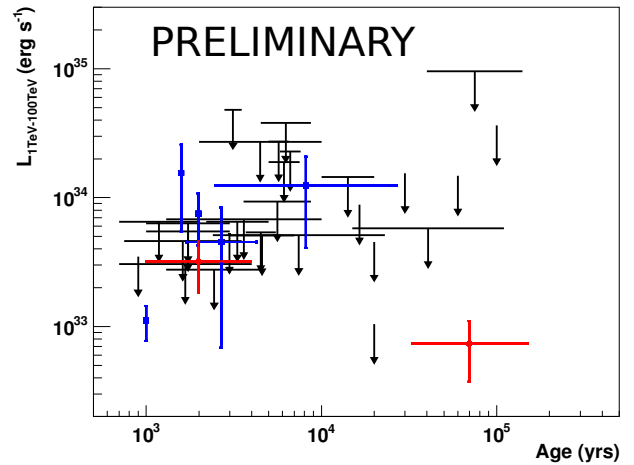}
          \caption{Upper limits on the integrated luminosity in the 1-100 TeV energy range versus the age for 29 SNRs with known distance and age of the Loose sample, assuming a spectral index of 2.3. Integrated luminosity  over 1-100 TeV of H.E.S.S. published shell SNRs (blue squares) and SNRs interacting with MC (red dots) are also shown on the plot.}
     	 \label{LumVsAge}
    \end{figure}

Figure \ref{LumVsAge} shows the upper limits on the integrated luminosity over the 1-100 TeV energy range versus the SNR age for 29 SNRs of the Loose sample, and assuming a spectral index of 2.3. The luminosity upper limit of about ten SNRs are at the same level as the integrated luminosity of 4 H.E.S.S. detected shell SNRs (blue squares), showing that the derived UL constrain the TeV SNR population.

\section{Conclusions}
In this work we present updated results on the study of undetected 
SNRs with H.E.S.S.. We employ a new selection theme that makes use of the signal
information we have at our disposal in the HGPS region. This results in a  analysis 
sample of 96-125 sources in total. 

We obtain integrated flux upper limit values above 1TeV using the HAP-TMVA method 
(cross-checked with the independent \textit{Model}-analysis) that allow us to test 
a commonly accepted theoretical estimate on the gamma ray luminosity from SNRs 
under the assumption of a hadronic-dominated emission via $\pi_0$-decay. The 
sample consists of 42 SNRs with a distance estimate and 30 SNRs where an additional 
age estimate is provided. 

Large uncertainties in the physical parameters of the studied SNRs limit the 
stringency of the obtained constraints on the acceleration efficiency $\theta$.

For the sample of SNRs with estimated distances, ULs on the source luminosities are
presented, constraining the Luminosity-Age parameter space. Luminosity values measured 
from detected SNRs agree with the obtained constraints.

This work shows first results on the continued SNR population study. Further 
results will be presented in a forthcoming publication.\\

\footnotesize{{\bf Acknowledgment:}{The support of the Namibian authorities and of the University of Namibia
in facilitating the construction and operation of H.E.S.S. is gratefully
acknowledged, as is the support by the German Ministry for Education and
Research (BMBF), the Max Planck Society, the German Research Foundation (DFG), 
the French Ministry for Research,
the CNRS-IN2P3 and the Astroparticle Interdisciplinary Programme of the
CNRS, the U.K. Science and Technology Facilities Council (STFC),
the IPNP of the Charles University, the Czech Science Foundation, the Polish 
Ministry of Science and  Higher Education, the South African Department of
Science and Technology and National Research Foundation, and by the
University of Namibia. We appreciate the excellent work of the technical
support staff in Berlin, Durham, Hamburg, Heidelberg, Palaiseau, Paris,
Saclay, and in Namibia in the construction and operation of the
equipment.}}

\end{document}